\documentclass{article}

\usepackage{arxiv}

\usepackage[utf8]{inputenc} 
\usepackage[T1]{fontenc}    
\usepackage{hyperref}       
\usepackage{url}            
\usepackage{booktabs}       
\usepackage{amsfonts}       
\usepackage{nicefrac}       
\usepackage{microtype}      
\usepackage{lipsum}		
\usepackage{graphicx}
\usepackage[numbers]{natbib}
\usepackage{doi}
\usepackage{xcolor}
\usepackage{caption}
\usepackage{amsmath}
\usepackage{subcaption}
\newcommand{\degree}{$^{\circ}$}

\title{Evaluation of Eye Tracking Signal Quality for Virtual Reality Applications: A
Case Study in the Meta Quest Pro}

\author{ 
    \href{https://orcid.org/0000-0002-7656-2662}{\includegraphics[scale=0.06]{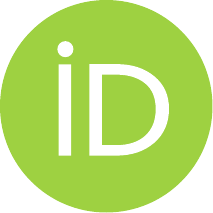}\hspace{1mm}Samantha Aziz} \\
	Department of Computer Science\\
	Texas State University\\
	San Marcos, TX 78666 \\
	\texttt{sda69@txstate.edu} \\
    \And
    \href{https://orcid.org/0000-0002-8088-9270}{\includegraphics[scale=0.06]{orcid.pdf}
    \hspace{1mm}Dillon J Lohr} \\
    Department of Computer Science\\
	Texas State University\\
	San Marcos, TX 78666 \\
	\texttt{djl70@txstate.edu} \\
    \And
    \href{https://orcid.org/0000-0002-6385-1035}{\includegraphics[scale=0.06]{orcid.pdf}
    \hspace{1mm}Lee Friedman} \\
    Department of Computer Science\\
	Texas State University\\
	San Marcos, TX 78666 \\
	\texttt{l\_f96@txstate.edu} \\
    \And
    \href{https://orcid.org/0000-0001-7890-8842}{\includegraphics[scale=0.06]{orcid.pdf}
    \hspace{1mm}Oleg Komogortsev} \\
	Department of Computer Science\\
	Texas State University\\
	San Marcos, TX 78666 \\
	\texttt{ok@txstate.edu} \\
}

\renewcommand{\shorttitle}{Evaluation of Eye Tracking Signal Quality for Virtual Reality in the Meta Quest Pro}

\hypersetup{
pdftitle={Evaluation of Eye Tracking Signal Quality for Virtual Reality Applications: A
Case Study in the Meta Quest Pro},
pdfsubject={cs.HC},
pdfauthor={Samantha Aziz, Dillon J Lohr, Lee Friedman, Oleg Komogortsev},
pdfkeywords={eye tracking, signal quality, data quality, quest pro, spatial accuracy, spatial precision},
}

\begin{document}
\maketitle

\begin{abstract}
We present an extensive, in-depth analysis of the eye tracking capabilities of the Meta Quest Pro virtual reality headset using a dataset of eye movement recordings collected from 78 participants. In addition to presenting classical signal quality metrics---spatial accuracy, spatial precision and linearity---in ideal settings, we also study the impact of background luminance and headset slippage on device performance.
We additionally present a user-centered analysis of eye tracking signal quality, where we highlight the potential differences in user experience as a function of device performance.
This work contributes to a growing understanding of eye tracking signal quality in virtual reality headsets, where the performance of applications such as gaze-based interaction, foveated rendering, and social gaze are directly dependent on the quality of eye tracking signal.
\end{abstract}


\label{sec:intro}
\section{Introduction}
Eye tracking technology is becoming increasingly common in consumer-grade computing platforms through its inclusion in virtual reality (VR) devices.
Eye tracking technology enables vital functionalities in VR environments, including gaze interaction techniques that replace conventional input modalities~\cite{Frenandes2023} and foveated rendering techniques that deliver vital power savings by selectively rendering parts of the screen based on the user's gaze~\cite{Guenter}.

The effective design and implementation of these eye tracking-based applications depends on the hardware and algorithms that collectively comprise the platform's gaze estimation pipeline.
From an application development standpoint, it is important to accurately gauge a device's eye tracking signal quality, so that applications may be designed relative to the platform's capabilities.
Both the accuracy and precision of gaze estimation constrain the design space for enabling a seamless user experience during gaze interaction, such as determining the minimum size of interactable elements and the minimum separation between such elements.
Similarly, the average accuracy and precision of gaze estimation determines important elements of foveated rendering that dictate the amount of power savings that can be attained, such as the size of the foveated region.

Because eye tracking is a relatively new interaction modality in VR, the best practices for designing eye tracking-based applications have not yet been fully articulated.
However, it is possible to apply principles from human factors engineering as a scaffold for developing these best practices.
For example, human factors engineering suggests that interaction-based platforms should be designed to accommodate the majority of users---a common rule is to accommodate all users that fall between the 5th and 95th percentile of a given operational capacity~\cite{guastello2023human}.
This principle can be practically applied for eye tracking by delegating spatial accuracy as the operational capacity in question.
Therefore, eye tracking-based applications should be designed to accommodate errors in gaze estimation so that the majority of users can use the application effectively. 

To enable such an operationalization, it is necessary to assess the eye tracking signal quality that is produced by a given platform. 
Eye tracking signal quality metrics are commonly reported in ways that are useful for eye tracking research, but the insights gained from this method are not necessarily useful for design practitioners who are looking to design applications for this novel interaction modality. 
Current signal quality analysis methods may not describe eye tracking signal quality in a manner that facilitates informed decision making when designing eye tracking-based applications in VR.
Namely, several studies report average eye tracking signal quality without sufficient regard for cases where eye tracking is likely to fail, and design decisions based on these metrics may not produce a suitable solution for any individual user~\cite{Privitera2019}. 
We argue it is not enough to merely design around the average accuracy exhibited by a device.
Users may be turned off from gaze interaction if it fails, even if failure is relatively rare.
As such, it is vital to account for the highest percentiles of gaze estimation error so that interface and interaction design can be tailored to minimize frustrating failures and ensure a seamless user experience during the vast majority of interactions.


This work addresses the need for eye tracking signal quality analysis that is informative for designing seamless gaze-based applications in VR.
In this work, we present an analysis of the eye tracking signal quality in the Meta Quest Pro, a consumer-grade VR device that includes eye tracking as a core functionality of the platform. 
We also introduce a new, intuitive framework that enables practical evaluation of eye tracking signals for applications in virtual reality, such as gaze interaction and foveated rendering.

The specific contributions of our work include:
\begin{itemize}
    \item An in-depth presentation of classical eye tracking signal quality metrics measured across a large population of 78 participants.
    \item A discussion on the implications of design decisions and environmental factors, such as variations in luminance and headset slippage, on eye tracking signal quality in the Meta Quest Pro.
    \item The introduction of user-centric eye tracking signal quality evaluation using user- and error- percentile tiers, which enables practical design of gaze-based applications in VR.
\end{itemize}
\label{sec:background}
\section{Background and Prior Work}

\subsection{Eye Tracking Applications in VR}
In addition to enabling vital power savings and improved rendering quality through foveated rendering, eye tracking technology has emerged as a promising avenue for improving interaction in VR platforms. 
Dwell time-based based gaze interaction enables users to complete selection-based tasks faster with fewer errors, especially in environments that span a large field of view~\cite{Blattgerste}.
Gaze input can also be combined with manual input via hand gestures~\cite{Pfeuffer2017} or controllers~\cite{Pfeuffer2020} to overcome problems associated with gaze-only input such as the Midas Touch~\cite{Mutasim}. 

Eye tracking can also be used to improve social interaction and collaborative efforts in VR environments by producing more lifelike virtual avatars.
As a form of non-verbal communication, gaze behavior provides social cues that play an important for effective face-to-face communication~\cite{jording2018social}. 
Using eye movement data to drive the gaze behavior of a user's virtual avatar can improve the perceived quality of communication between human parties in VR~\cite{Garau, Steptoe2008, Bente}.
The appearance of the eye contributes greatly to a user's overall impression of how lifelike an avatar appears to be~\cite{Looser}; incorporating motion-captured gaze behavior into a virtual avatar's expressions can make virtual avatars appear more expressive and can enable more effective, naturalistic interaction.
Gaze-driven avatar animations also enables two or more parties to mutually coordinate their attention in a shared virtual environment through their virtual avatars, which improves can improve the outcomes of collaborative tasks~\cite{Steptoe2009}.

\subsection{Eye Tracking Signal Quality Analysis}
The effectiveness of eye tracking-based applications depends on the platform's eye tracking signal quality, or the degree to which a user's eye movement can be reliably and accurately captured by an eye tracking device. 
It is not sufficient to rely on descriptions of eye tracking signal quality metrics reported by device manufacturers because these metrics cannot reliably be replicated in practical settings; discrepancies between the manufacturer-reported performance metrics and empirical results are often observed, even in controlled laboratory setups~\cite{Zhang2011, Aziz2022, Komogortsev}.
The eye tracking signal quality that can be captured by a device can vary substantially based on experimental factors such as recording admin experience~\cite{holmqvist2011eye}, user characteristics such as race, eye color, and and eyelid shape~\cite{Blignaut2014, Holmqvist2017}, and user behavior such as head movement~\cite{Niehorster2018}.
It is therefore important to characterize eye tracking signal quality using data gathered empirically in a variety of conditions, and to report on those conditions when they may affect the quality of the data that is collected~\cite{Dunn2023}.

While we are not the first to evaluate the Meta Quest Pro's eye tracking capabilities, our study expands the characterization of the device's performance in several dimensions.
Prior work in assessing signal quality in the Quest Pro~\cite{Wei2023} presented a preliminary evaluation of the device's eye tracking signal quality by measuring the spatial accuracy and spatial precision across 12 people in both constrained and unconstrained settings.
We expand on these preliminary results by reporting the same classical eye tracking signal quality metrics across a significantly larger population of 78 participants. 
We also present a more nuanced perspective on spatial accuracy through our exploration of how it systematically varies across the field of view, which can have significant design implications.
We also explore the robustness of the device to two environmental factors that are uniquely challenging in VR: background luminance and headset slippage.


\subsubsection{Effect of Luminance on Eye Tracking Signal Quality}
The pupil's sensitivity to environmental changes has historically been an important consideration for video-based eye tracking systems, as changes in ambient lighting can cause the pupil to constrict and dilate in ways that are not conducive to consistent gaze estimation across environmental conditions~\cite{Hutton2019, Nystrom2016}.
Environmental luminance, and by extension pupil size, has been observed to impact the spatial accuracy of gaze estimation, and this effect varies by the platform's underlying gaze estimation pipeline.
For example,~\cite{Holmqvist2012} observe deterioration in eye tracking signal quality across 11 video-oculography based eye trackers when eye tracking data is collected in a dark environment.
While details about the gaze estimation pipeline in the Meta Quest Pro remain proprietary knowledge, it is still meaningful to assess the effect of luminance (and, by extension, changes in pupil diameter) on the platform's ability to accurately estimate gaze. 

Analysis of how luminance impacts eye tracking changes when considering eye tracking in VR platforms, as VR headsets are designed to shield users from ambient light.
As a result, luminance is controlled by the artificial environment created within the headset, and the magnitude of interference from external sources of light is relatively low compared to more conventional (i.e., remote) eye tracking environments.
Rather than considering ways to mitigate the interference from external sources of light, VR environments are concerned with the designing artificial luminance conditions that balance high-quality eye tracking and user comfort.
Although lower luminance conditions are associated with improved levels user comfort over prolonged exposure to near-eye displays~\cite{Hirzle, Sang, Erickson}, users have reported that they favor brighter luminance conditions in VR environments because they are perceived as more realistic and immersive than low-luminance conditions~\cite{Matsuda}.
As a result, both high- and low-luminance conditions are realistic choices for implementation in VR environments. 
Although factors like cognitive load and emotion impact pupil size~\cite{wang2011pupil}, we focus on investigating the effects of luminance because it is a design component that practitioners have the most control over when attempting to reduce the potential influence of pupillary activity on eye tracking signal quality in VR. 


\subsubsection{Slippage in Head-Mounted Eye Tracking Devices}
Headset slippage (or simply ``slippage'') is an issue that has emerged with the proliferation of wearable eye trackers, including those embedded in VR devices.
Slippage occurs when the headset's position shifts on the user's face, which changes the position of the eye tracking sensors relative to the eye at the time the device was calibrated and results in a deterioration of eye tracking signal quality.
The degree to which slippage may occur and the resulting impact on eye tracking signal quality may be affected by a number of factors, such as the quality of the straps used to secure the headset to the user’s head, the robustness of the hardware to repositioning, and the degree to which an individual user may induce slippage through regular activity like speaking and making facial expressions. 
Nevertheless, headset slippage has been observe to negatively impact the validity and quality of eye tracking signals collected in wearable eye trackers~\cite{Niehorster2018, Niehorster2020}.
We contribute to a growing body of knowledge in this direction by reporting the effect that slippage may have on the Meta Quest Pro's performance to inform the development of design practices that can compensate for slippage when it inevitably occurs. 



\label{sec:method}
\section{Methodology}
\subsection{Participants}
We recruited 78 college-aged participants (28 male, 50 female, mean age 19.5 years, age range 18-28) from the student population of Texas State University.
All participants had normal or corrected-to-normal vision; 13 participants wore glasses and 14 participants wore contact lenses while completing this experiment. 
Three glasses-wearing participants removed them for this experiment, due to issues fitting their glasses frames in the device. 
All participants reported being able to see the stimulus without requiring additional vision correction. 
Participants were seated for the duration of the experiment and used a chinrest to minimize extraneous head movement.

\subsection{Stimulus and Apparatus}
Eye movement recordings were captured on an eye tracker embedded within a Meta Quest Pro VR headset.
This eye tracking device is included as standard hardware on the Quest Pro, and captures positional eye movement data at a sampling rate up to 90 Hz through Meta's public eye tracking API for Unity~\cite{meta-api}.
For this investigation, we measure eye tracking signal quality on a cyclopean ray produced by computing the average gaze position between the two eyes.

Prior to data collection, we ensure that the headset was optimally positioned on each participant's head using the positioning guidance system that is built into the platform. 
Once the device was optimally secured on the participant's head, the eye tracking device was calibrated using the device's built-in 9-point grid calibration procedure. 
Calibration took place once at the beginning of data collection process, and was generally not repeated unless the participant removed the headset between tasks.

Participants completed a series of eye tracking tasks that were designed to elicit eye movements that are useful for signal quality evaluation. 
Prior to each task, the headset displayed instructions on how to complete each task.
Participants were instructed to initiate the task once they understood the instructions.

\begin{figure}
    \centering
    \includegraphics[width=0.43\linewidth]{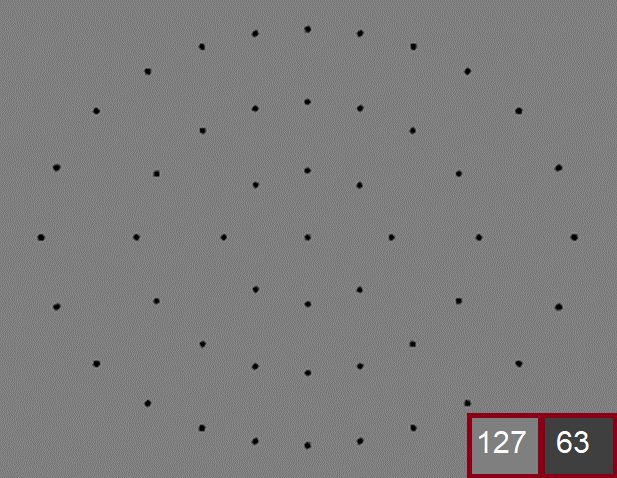}
    \caption{The orientation of the targets during the random saccades tasks, including the colors employed for RAN 127 and RAN 63.}
    \label{fig:ran-task}
\end{figure}

Participants completed a battery of five eye tracking tasks.
For this experiment, we highlight three tasks that facilitate analyses of eye tracking signal quality in the presence of luminance variation and headset slippage.
These tasks, along with the abbreviations that we will employ throughout this manuscript, are described in the order in which they were presented during data collection:

\begin{enumerate}
    \item \textbf{Random Saccades, Bright Background (RAN 127):} Participants followed a small black dot as it jumped to random positions on the screen along an elliptical grid spanning $\pm$25 degrees of visual angle (dva, \degree) horizontally and $\pm$20 dva vertically at a distance of 1 meter. Forty-nine fixation targets were uniformly distributed in this range. At each position, the stimulus initially appeared as a dot with an apparent diameter of 2.3 dva, then shrank to the size of 0.57 dva over the course of 1 second, then remained at that size for 1 additional second, for a total of 2 seconds at each position. The stimulus was presented against a high-luminance background whose color corresponds to a light gray (127, 127, 127) on the RGB scale ranging from [0,255]. The target was head-locked during this task, meaning that the position of the target relative to the center of the headset remained invariant. Figure~\ref{fig:ran-task} includes a visualization of this task, including the orientation of the stimuli.
    \item \textbf{Brow Task (BROW):} Inspired by an eye tracking task introduced in prior studies of headset slippage~\cite{Niehorster2020}, this task induces headset slippage via eyebrow movement. Participants gazed at a small black sphere with an apparent diameter of approximately 0.5 dva at a distance of 1 meter in the center of their field of view. The headset was programmed to intermittently play a sound; participants would raise their eyebrows and hold them in the raised position for a duration of 2 seconds. Then, the headset would play a second sound, which prompted participants to relax their eyebrows for 5 seconds. This ``brow-up-brow-down'' procedure was repeated 5 times, and participants gazed at the target in the headset for the duration of the experiment. The stimulus was world-locked during this task, meaning that subjects could continue looking straight ahead at the target, even when the headset physically shifted on their head.
    \item \textbf{Random Saccades, Darker Background (RAN 63):} This task is identical to RAN 127, but the background color was set to a darker shade of gray that corresponds to (63, 63, 63) on the RGB scale ranging from [0,255].
\end{enumerate}


\subsection{Signal Quality Evaluation}
\subsubsection{Data Processing and Signal Quality Evaluation}
While eye tracking signal quality analysis is far from being standardized in the field, such analyses typically present evaluations of eye tracking signal quality from the perspective of classical metrics, including spatial accuracy and spatial precision.

We only calculate signal quality metrics during stable fixation periods, when we assume that the eye is in a relatively stationary state.
Rather than developing a fine-tuned event classification algorithm to classify periods of stable fixation in the gaze data, we identify stable fixation periods relative to the timing of target movement.
While this approach does not guarantee that saccadic movement is fully eliminated from the portions of the signals that we employ for analysis, it eliminates the uncertainty that could arise from sub-optimal event classification.

For the random saccades tasks (RAN 127 and RAN 63), we discard the first 1 second of each target presentation period to account for both saccade latency and general instability, and employ the subsequent 0.5 seconds of eye movement data for measurement.
During the brow task, we employ a similar strategy during the 2-second periods when the participant's brows are raised---we discard the first 1 second of each period and employ the next 0.5 seconds for analysis.
During the 5-second periods when the participant's brows are relaxed, we skip the first 4.5 seconds during each period and employ the last 0.5 seconds for measurement.
We choose these periods of time to minimize the temporal gap between pairs of ``brow up'' and ``brow down'' periods. 

Spatial accuracy is calculated as the angular distance between the target position and the participant's 3D gaze vector during stable fixation periods (Equation~\ref{eqn:accuracy}),

\begin{equation}
\label{eqn:accuracy}
    \theta = \frac{180}{\pi} \arccos\left(\frac{g \cdot t}{\lVert g \rVert \lVert t \rVert}\right),
\end{equation}
where $g$ is the gaze vector, $t$ is the eye-to-target vector, and $\theta$ is angular distance in degrees of visual angle.
Spatial precision is calculated as the sample-to-sample root mean square precision.



We examine the distribution of observed values across the user population for both metrics.
Although we include measurements of the mean values we observe to facilitate comparisons with prior work, we argue that the mean is a particularly reliable way to describe distributions of spatial accuracy and precision, as they tend to be highly skewed distributions and frequently contain extreme outliers.
Instead, we will rely on descriptions of user- and error- based percentiles to better capture the typical values observed within users and variability across the population.


\subsubsection{U Percentiles: User-Based Evaluation Across a Population}
To facilitate evaluations of eye tracking signal quality that is informative for user-centric design, we introduce two evaluation techniques that places users at the center of our signal quality assessment.
The first of these techniques is user percentiles, which are denoted with ``U'' through this manuscript.
User percentiles can be used to examine the variability in signal quality across members of the population.
A user whose average spatial accuracy metrics represent the median spatial accuracy across the entire user population would be denoted as the U50 user. 
This is useful because it allows design practitioners to describe the behavior of an average user across the range of eye tracking signal quality levels that may be experienced during normal operation of an eye tracking device. 
If a design practitioner has designed an eye tracking application with an ideal spatial accuracy in mind---derived, perhaps, from a theoretical framework meant to produce accuracy requirements for various interaction techniques~\cite{Swafford2016,Zhang2007,Feit2017}---then they could leverage user percentiles to observe what percentage of the user population meets that ideal.
Framing signal quality evaluation this way allows designers to assess whether their application design would be usable by a sufficiently large percentage of users across the operational spectrum of the eye tracking platform. 

\subsubsection{E Percentiles: Error-Based Evaluation Within Users}
The second evaluation technique is individual error percentiles, which are denoted with ``E'' throughout this manuscript.
Error percentiles describe the distribution of signal quality metrics across all stable fixation periods within an individual user's eye movement recording. 
For example, the E50 spatial accuracy for a single user represents the median gaze error that was measured in their gaze data, E75 represents the 75th percentile, and so on. 
While user percentiles assess population coverage, error percentiles assess coverage for specific users.
This is vital for assessing whether they would have a consistent gaze interaction experience.  

User percentile and error percentile metrics can be combined to indicate performance numbers for a specific user representing a notable average percentile of a given signal quality metric across the sample. 
In this way, ``U50|E95 spatial accuracy'' is shorthand for ``the 95th percentile of spatial accuracy error for the user representing the 50th percentile of eye tracking signal quality across the sample.'' 
Taken together, user percentiles and error percentiles enable informed decisions to be made for interface and interaction design to ensure an useful experience for a vast majority of users the majority of the time.

\label{sec:results}
\section{Results}
\subsection{Spatial Accuracy}
\begin{figure}
    \centering
    \includegraphics[width=0.49\linewidth]{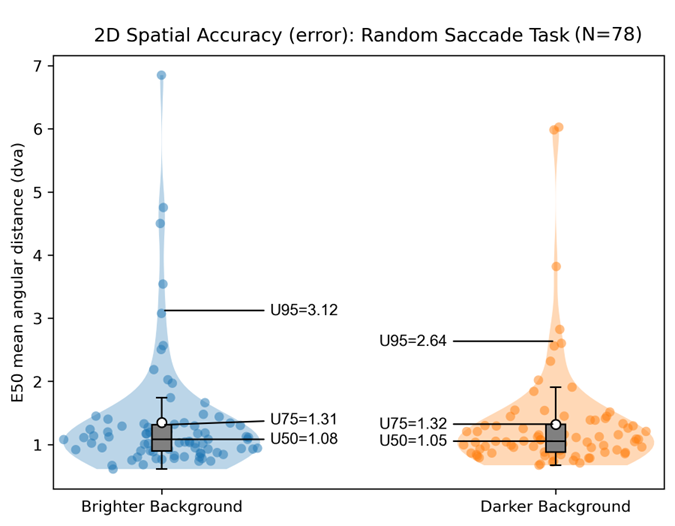}
        \includegraphics[width=0.49\linewidth]{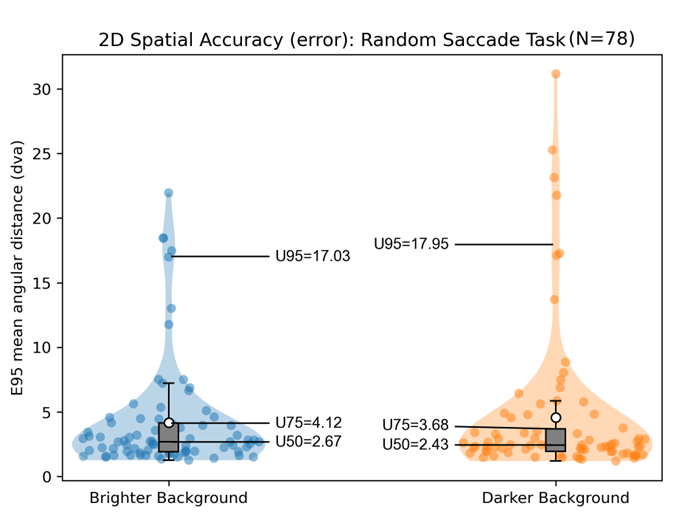}
    \caption{The distribution of spatial accuracy calculated across both luminance conditions. Each data point represents E50 (left) and E95 (right) mean spatial accuracy from a user. The white circle represents the mean spatial accuracy across the entire population.}
    \label{fig:accuracy-luminance}
\end{figure}

Spatial accuracy describes the difference between the user's gaze position and that which is reported by an eye tracking device.
We report the spatial accuracy observed in the Meta Quest Pro measured across both luminance conditions during the random saccades tasks in Figure~\ref{fig:accuracy-luminance}.
We include the distribution of E50 spatial accuracies to facilitate comparisons with conventional descriptions of eye tracking signal quality, and to describe the average spatial accuracy of users during normal device operation.
Figure~\ref{fig:accuracy-luminance} (left) shows the distribution of spatial accuracy across the population, where each data point represents the median (E50) spatial accuracy aggregated across each user's eye tracking signal.
The distribution of E95 spatial accuracy metrics shown in Figure~\ref{fig:accuracy-luminance} (right) represents the distribution of the worst-case spatial accuracy produced by each user during the random saccades tasks. 
Unsurprisingly, we observe a significant performance discrepancy between E50 and E95 distributions.
While U50|E50 and U75|E50---both depicted in Figure~\ref{fig:accuracy-luminance} (left)---are close to one another, the separation between U50|E95 and U75|E95 feature notably more separation.
While the majority of the data points in the E95 are clustered toward the lower end of the range, there are a few extreme spatial accuracy values that pull the values of U75|E95 and U95|E95 away from the metrics measured on the majority of users. 
It appears that the spatial accuracy produced by the Meta Quest Pro is robust to background luminance, as both the E50 and E95 distributions of spatial accuracy are consistent between luminance settings.

\begin{figure}
    \centering
    \includegraphics[width=0.49\linewidth]{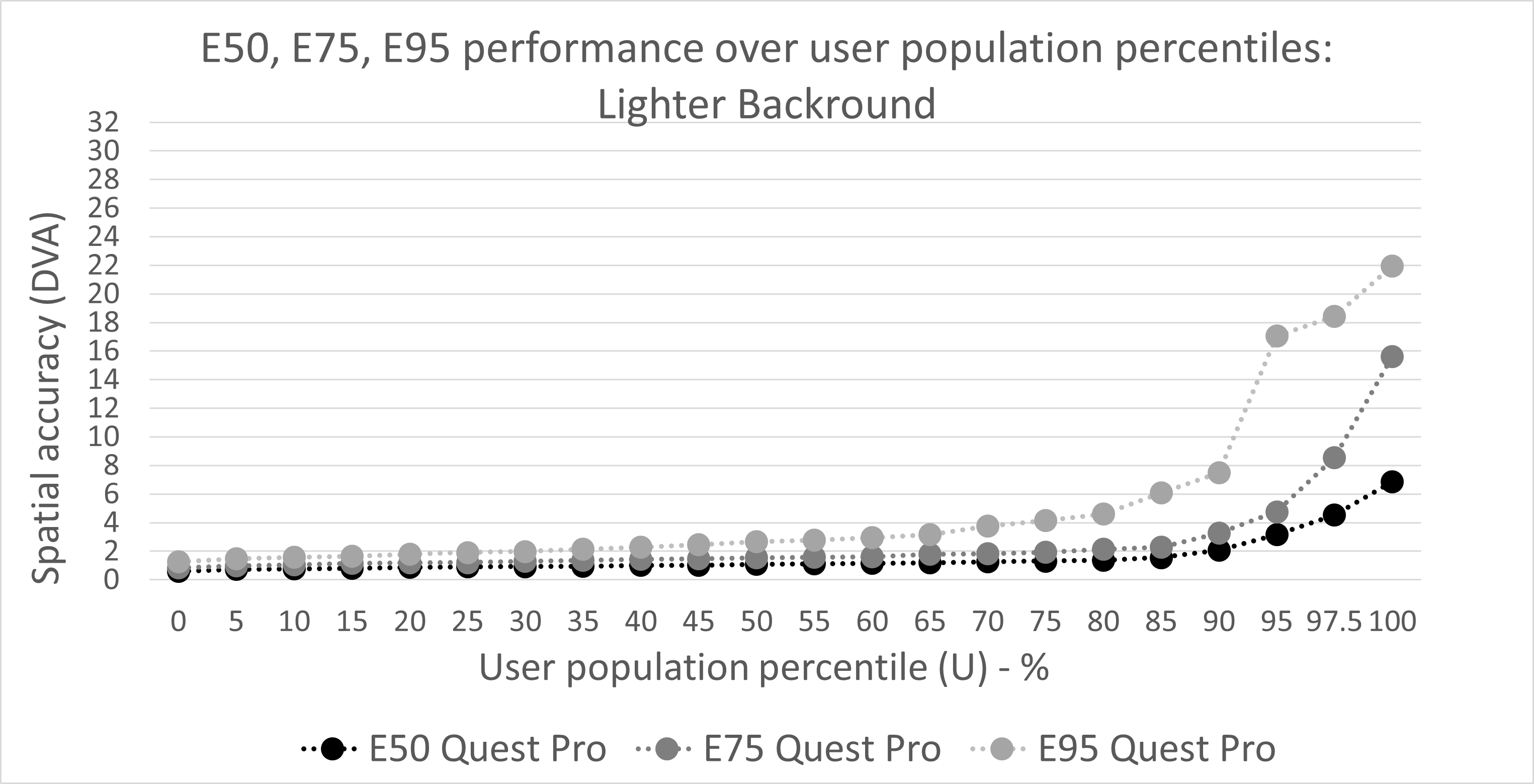}
    \includegraphics[width=0.49\linewidth]{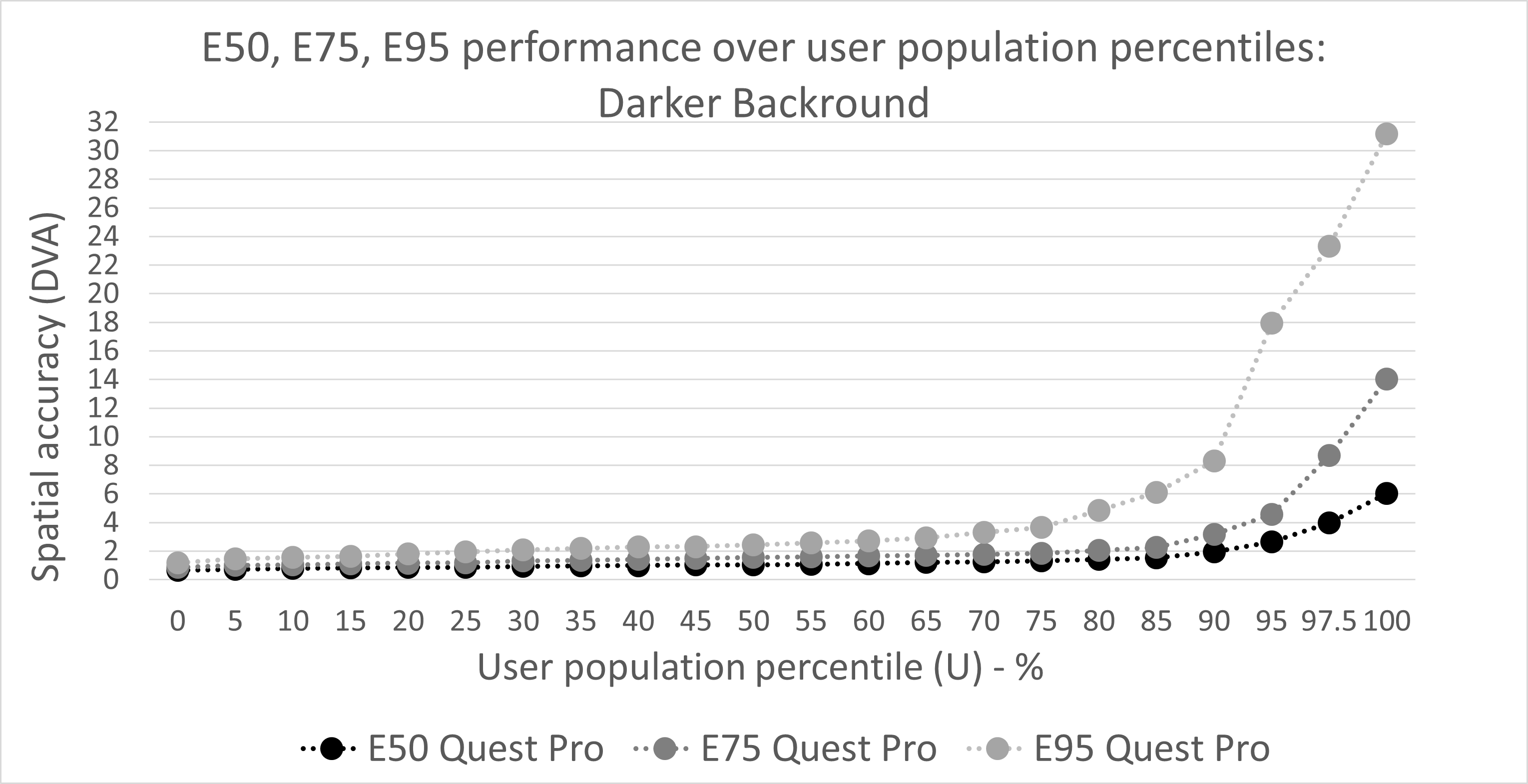}
    \caption{The spatial accuracy requirements to accommodate different proportions of the user population in both the bright (left) and dark (right) background random saccades tasks. The E50, E75, and E95 metrics are presented for all user percentiles shown here.}
    \label{fig:E-performance-luminance}
\end{figure}

Figure~\ref{fig:E-performance-luminance} further explores the relationship between user percentiles and spatial accuracy by describing the E50, E75, and E95 spatial accuracy observed at each user percentile tier.
This visualization enables an assessment of the extent to which a specific eye tracking application can be effective for a certain user population, considering its spatial accuracy requirements.
For example, a system-critical eye tracking application that requires an average spatial accuracy of 6 dva across 95\% of an individual user's interaction experience can serve 85\% of users against a lighter background and 80\% of users against a darker background. 

\subsection{Spatial Precision}
\begin{figure}
    \centering
    \includegraphics[width=0.49\linewidth]{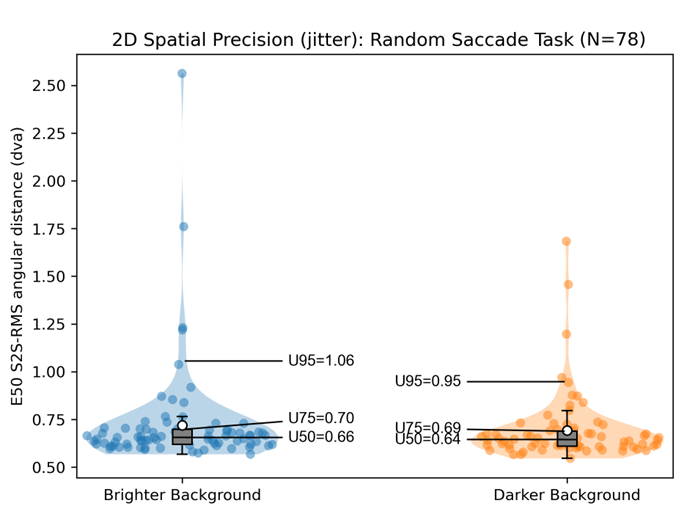}
    \includegraphics[width=0.49\linewidth]{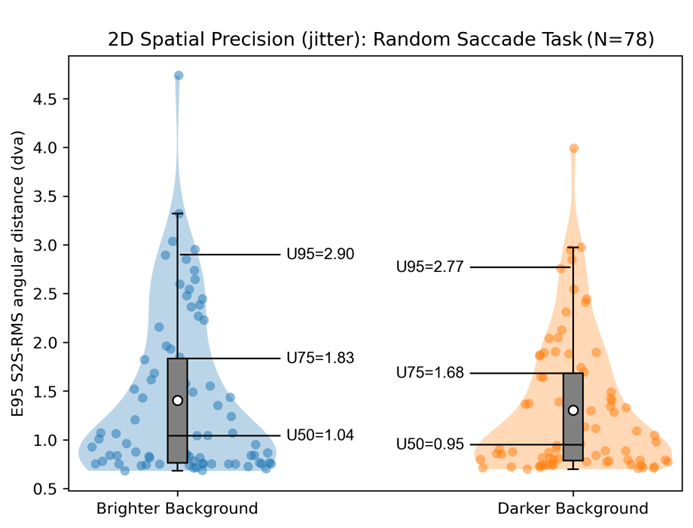}
    \caption{The distribution of sample-to-sample (S2S) RMS spatial precision across both luminance conditions. Data points represent E50 (left) and E95 (right) of mean spatial precision from a user. The white circle denotes the mean spatial precision across the population.}
    \label{fig:precision-luminance}
\end{figure}

Spatial precision describes the level of variability or instability in the eye movement signals produced by the device when a user is fixating on a single point in their field of view.
High levels of sample-to-sample instability may necessitate the use of real-time filtering techniques to smooth out the eye movement signal, which introduces additional latency and processing overhead to the gaze estimation pipeline.
Figure~\ref{fig:precision-luminance} shows the distribution of the E50 and and E95 RMS spatial precision achieved across the population.
Similar to our analysis of spatial accuracy, we include both the distribution of E50 and E95 metrics to respectively describe the average and worst-case spatial precision that one could expect to achieve on the device across both luminance conditions. 

The distribution of E50 spatial precision is shown in Figure~\ref{fig:precision-luminance}, (left); each data point in this figure represents the median spatial precision observed for the user in both the RAN127 (brighter background) and the RAN63 (darker background) tasks.
The average spatial precision measured across users tends to not exceed 1 dva, and features relatively low variability with few outliers. 
It does not appear that different luminance conditions affect average spatial precision.

While the average spatial precision exhibited across users tends to feature little variation, much larger variation between users arises when examining the worst-case spatial precision metrics across the population.
The distribution of E95 spatial precision is shown in Figure~\ref{fig:precision-luminance}, (right), where each data point represents a user's E95 spatial precision measured in both luminance settings.
This distribution shows a much wider gap in the worst-case metrics between users in the U50 population tier and users at the U75 and U95 population tiers. 
Unlike the E50 distribution, a subgroup of users with extremely poor spatial precision clearly emerges in the E95 distribution.

These findings emphasize the significance of assessing the signal quality metrics under the worst-case conditions. 
If an eye tracking application were designed to support the users based on the median E50 metrics observed in this population (U50|E50 = 1.08 for RAN127, 1.05 for RAN63), it would not perform effectively for more than half of the users when considering their corresponding E95 metrics (U50|E95 = 2.67 for RAN127, 2.43 for RAN63).
Considering that these metrics were collected within the duration of a relatively short eye tracking task, it is reasonable to speculate that the consequences of this design flaw would occur frequently.

\subsection{Impact of Slippage on Spatial Accuracy and Spatial Precision}

Inspired by the characterization of headset slippage presented by~\cite{Niehorster2020}, we measure the impact of headset slippage on spatial accuracy through an eye tracking task that prompts users to raise and lower their eyebrows as they fixate on a stationary target.
The brow movement done during this task shifts the device on a participant's head.
After the series of five brow movements that are performed, there is a high
likelihood that the device is at a different position compared to where it was at the beginning of the task.
We assess the impact of this movement on spatial accuracy.

\begin{figure}
     \begin{subfigure}[b]{0.49\linewidth}
         \centering
        \includegraphics[width=\linewidth]{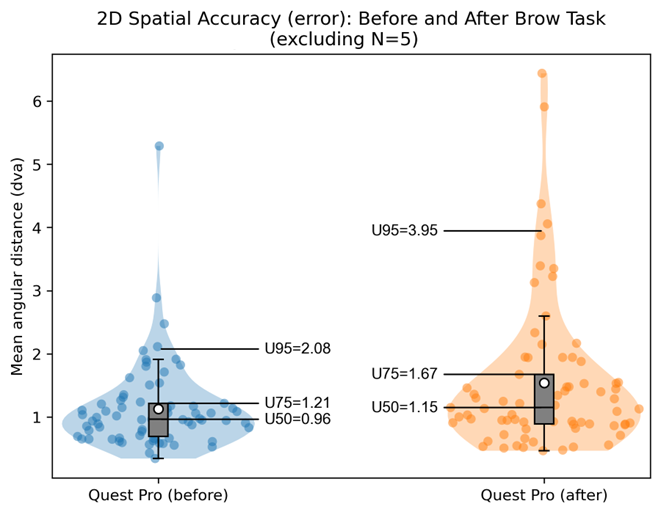}
        \caption{Before versus after brow task.}
    \label{fig:accuracy-slippage-before-after}
     \end{subfigure}
     \begin{subfigure}[b]{0.49\linewidth}
         \centering
        \includegraphics[width=\linewidth]{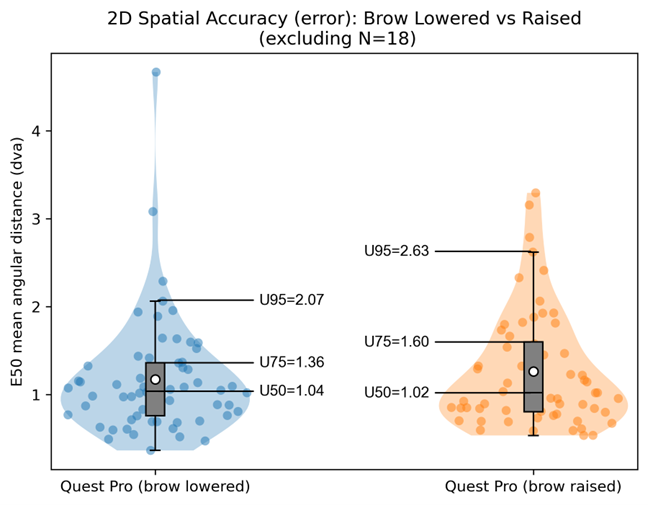}
        \caption{Brow lowered versus brow raised during the brow task.}
        \label{fig:accuracy-slippage-intra-task}
     \end{subfigure}
    \hfill
    \caption{Violin plots comparing spatial accuracy before and after the brow task (\ref{fig:accuracy-slippage-before-after}) and during the brow task (\ref{fig:accuracy-slippage-intra-task}). Each data point represents a single mean from each user. The white circle represents mean spatial accuracy across the population.}
    \label{fig:slippage}
\end{figure}

Figure~\ref{fig:accuracy-slippage-before-after} shows the change in spatial accuracy measured before and after participants completed the brow task. 
For both of the analyses, we excluded participants from analysis if we failed to collect data during any of the fixation periods during the task or if any of these fixation periods exhibited spatial error greater than 10 dva.
Five participants were excluded based on these criteria.
While completing this task tends to produce relatively low degradation in spatial accuracy across the E50 and E75 population overall, it clearly exacerbates problems with device performance at the highest population tiers.
We observe that the E50 user percentile does not experience a significant degradation in spatial accuracy after completing the brow movement task (U50|E50 is equal to 0.96 dva and 1.15 dva, respectively).
However, many participants along higher user percentile tiers experienced significant spatial accuracy degradation after completing the brow task, indicating that headset slippage reduces the accuracy and reliability of the gaze data captured in the U75+ the population.

Figure~\ref{fig:accuracy-slippage-intra-task} shows the distribution of spatial accuracy as participants completed the brow task.
Spatial accuracy is computed during each of 5 ``brow down'' periods and each of the 5 ``brow up'' periods for each participant; the distribution of the E50 spatial accuracy metrics observed across the population are then compared between the time when the brows are raised versus when the brows are lowered.
Eighteen of the 78 subjects were excluded from analysis, which illustrates the effect that headset slippage can have in the short term.
In general, we observe worse spatial accuracy during the ``brow up'' period.
Inducing headset movement appears to degrade average eye tracking performance across the population, and tends to produce more users with relatively higher spatial error.
However, when the headset is shifted back into the position it was originally calibrated in (i.e., when the user's brows are relaxed), the device is generally robust to minor variations in headset position as a result of the initial headset slippage. 



\label{sec:linearity}
\subsection{Linearity}
\begin{figure}
    \centering
    \includegraphics[width=.85\linewidth]{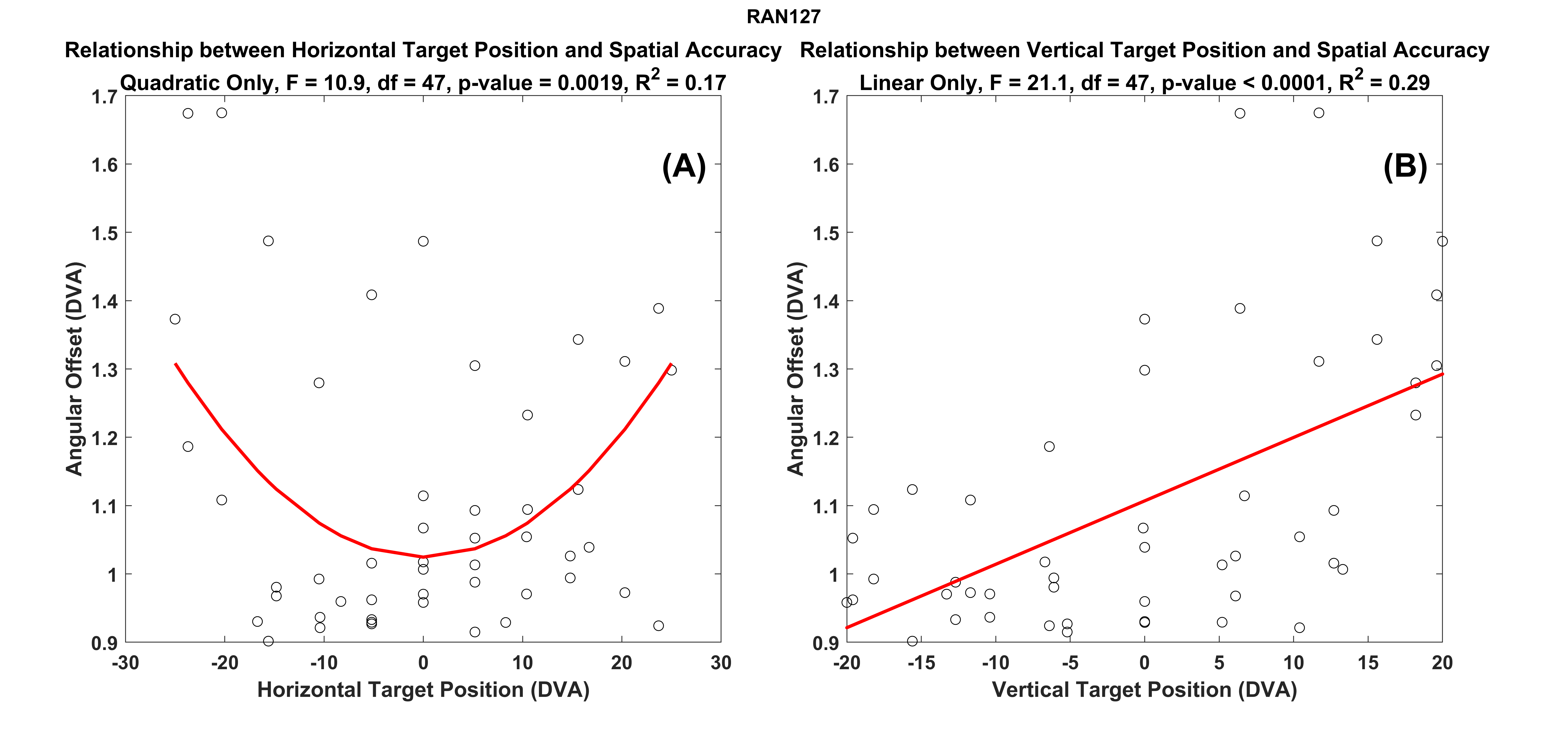}
    \caption{The relationship between spatial accuracy and and target position for the bright background random saccade task. (A) Horizontal target position versus spatial accuracy. (B) Vertical target position versus spatial accuracy.}
    \label{fig:1}
\end{figure}

\begin{figure}
    \centering
    \includegraphics[width=.85\linewidth]{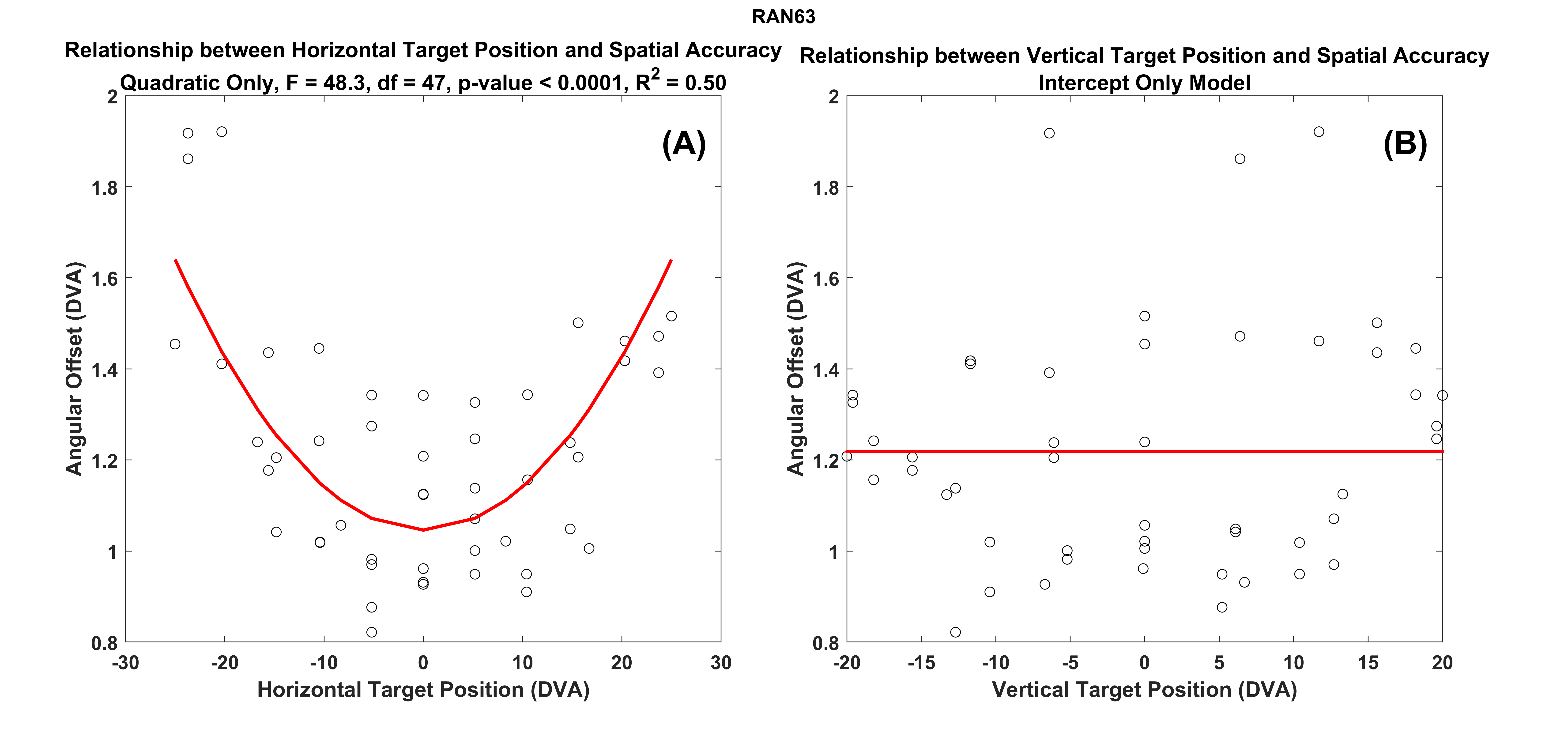}
    \caption{The relationship between spatial accuracy and and target position for the dark background random saccade task. (A) Horizontal target position versus spatial accuracy. (B) Vertical target position versus spatial accuracy.}
    \label{fig:2}
\end{figure}

Linearity describes the extent to which spatial accuracy depends on the position of the target in space.
Eye tracking signal quality is known to vary across the field of view, often degrading toward the periphery~\cite{Hornof2002, Aziz2022, lohr2019}.
We test the relationship between target position and spatial accuracy in Figures~\ref{fig:1} and~\ref{fig:2}.
Figure~\ref{fig:1} illustrates the relationship between spatial accuracy and target position for RAN 127, the bright background random saccades task.
Figure~\ref{fig:2} illustrates the relationship between spatial accuracy and target position for RAN 63, the dark background random saccade task.
These relationships were tested using a stepwise multiple linear regression analysis (second order polynomial fits) with angular offset as the dependent variable and target position and target position\textsuperscript{2} (squared) as independent variables.

For the horizontal analysis for RAN127 (Figure~\ref{fig:1} (A)), the stepwise regression only used the quadratic component (position\textsuperscript{2}) predictor. 
The result is a parabolic shape relationship, where spatial accuracy is worse when the horizontal target was in the left or right periphery of the visual field.
For the vertical analysis (Figure~\ref{fig:1} (B)), the stepwise regression only used the linear (position) predictor.  
Spatial error is lower when the target is below the center of the display and higher when the target is above the center of the display.

For the analysis of the impact on the target's horizontal position on spatial accuracy for RAN63 (Figure~\ref{fig:2} (A)), the stepwise regression only used the position\textsuperscript{2} predictor.
The result is also a parabolic shape, with worse spatial accuracy at the left or right periphery. 
For the vertical analysis (Figure~\ref{fig:2} (B)), the stepwise regression did not use either the linear or the quadratic predictor.
In this case, there was no relationship between spatial error and vertical target position.

\begin{figure}
     \begin{subfigure}[b]{0.33\linewidth}
         \centering
         \includegraphics[width=\linewidth]{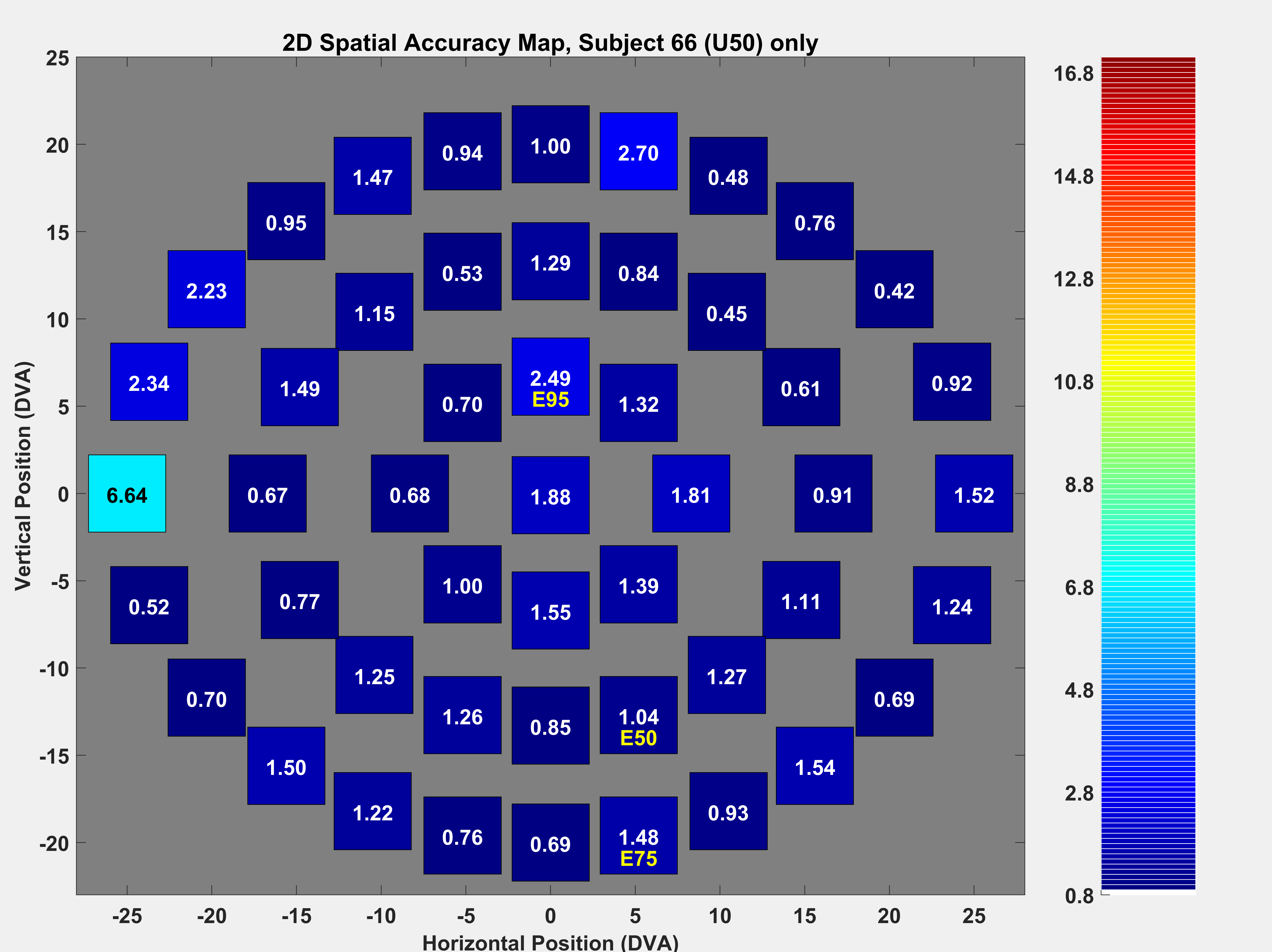}
         \caption{Subject 66, a typical subject.}
         \label{fig:U50-RAN63}
     \end{subfigure}
     \begin{subfigure}[b]{0.33\linewidth}
         \centering
         \includegraphics[width=\linewidth]{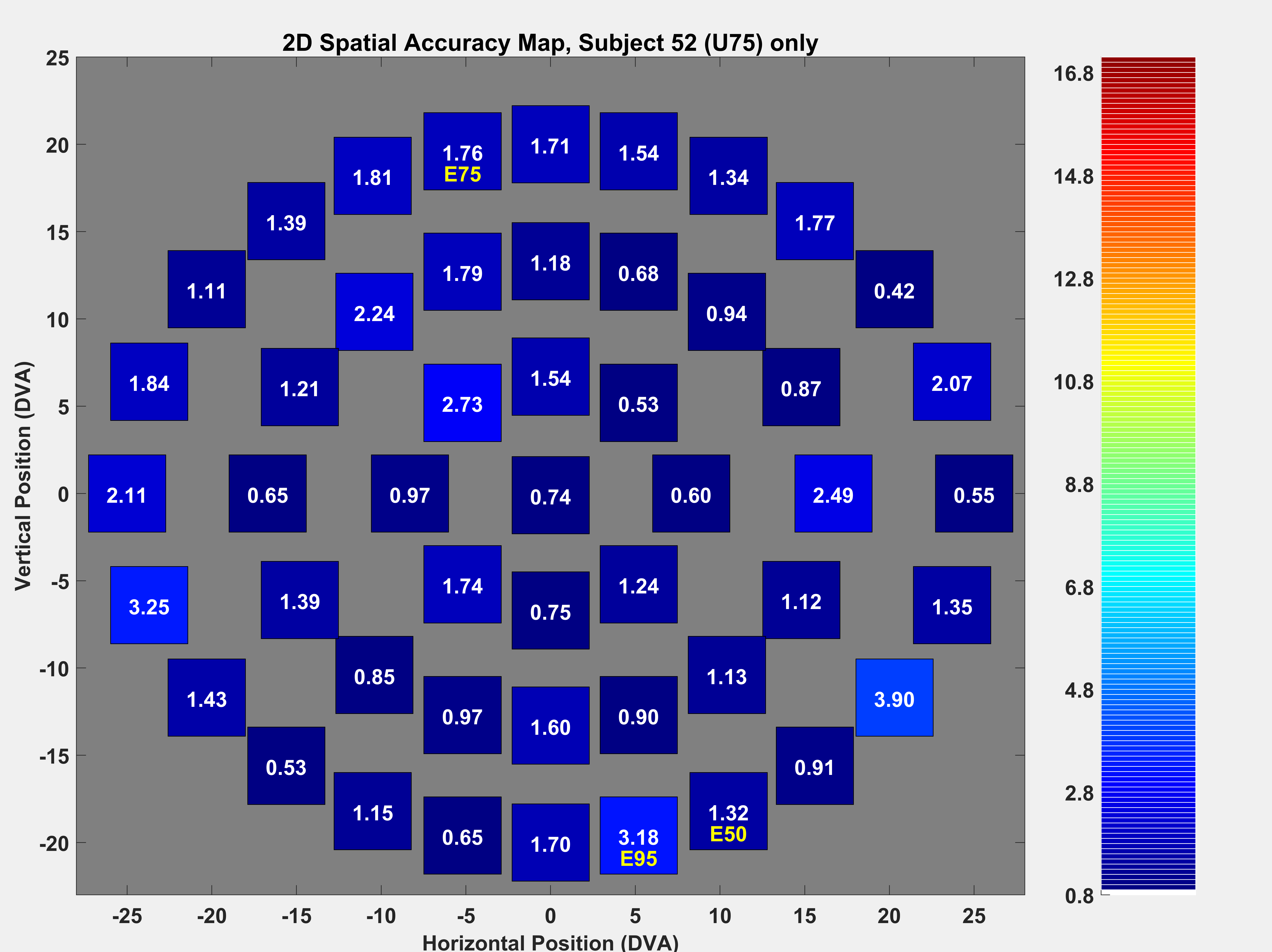}
         \caption{Subject 52, a moderate-error subject.}
         \label{fig:U75-RAN63}
     \end{subfigure}
     \begin{subfigure}[b]{0.33\linewidth}
         \centering
         \includegraphics[width=\linewidth]{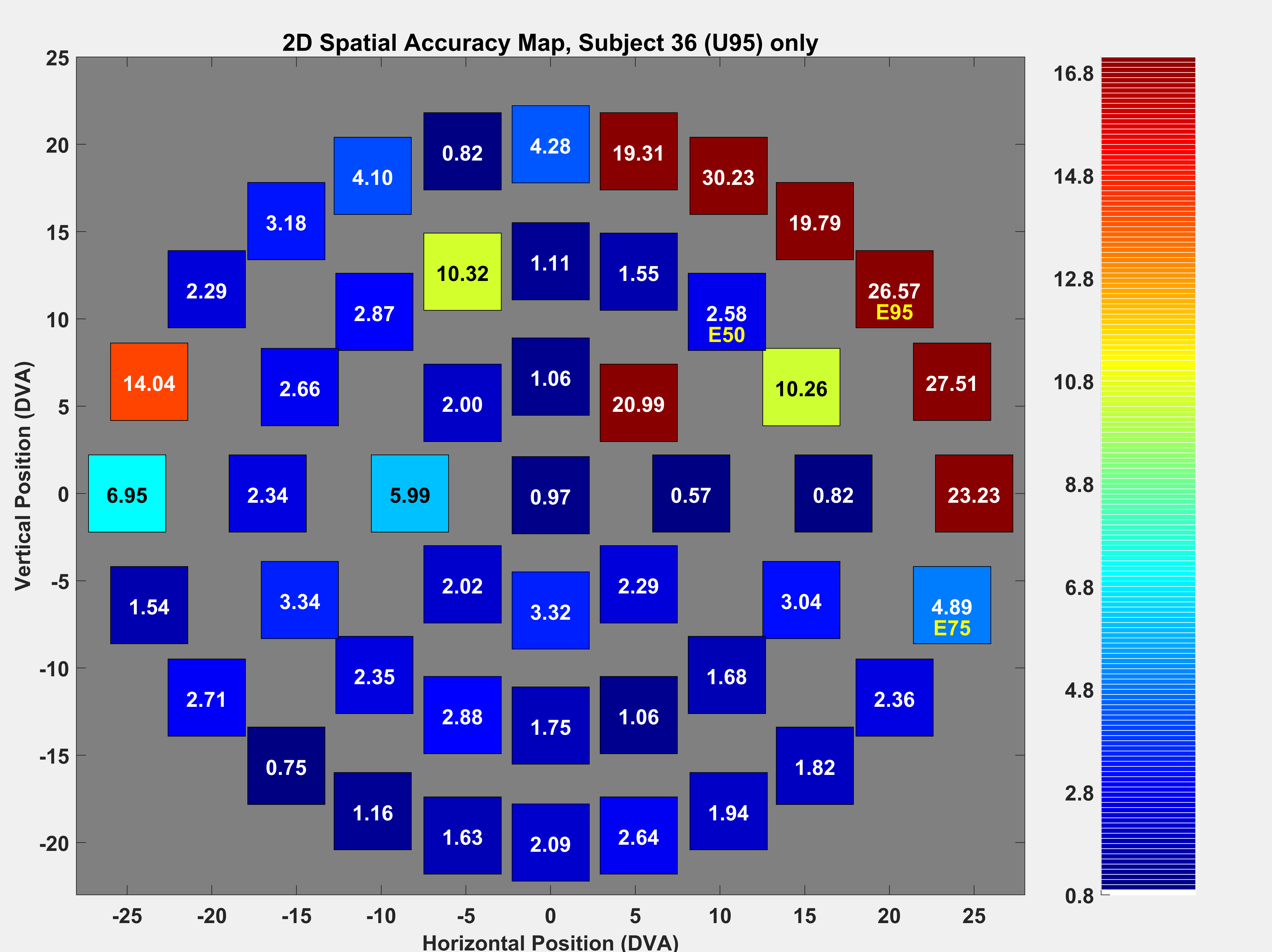}
         \caption{Subject 36, a high-error subject.}
         \label{fig:U95-RAN63}
     \end{subfigure}
    \hfill
    \caption{A two-dimensional representation of the spatial accuracy across the field of view, exemplified using the U50, U75, and U95 participants for the RAN63 and RAN127 tasks, respectively.}
    \label{fig:2d-spatial-accuracy}
\end{figure}

These findings describe the relationship between target position and spatial accuracy across the entire dataset of eye movement signals.
For an analysis on an individual level, the 2D spatial accuracy maps presented in Figure~\ref{fig:2d-spatial-accuracy} can provide useful examples of how this relationship may manifest in individual users. 
Figure~\ref{fig:2d-spatial-accuracy} displays the variation of spatial accuracy across space for a representative sample of users.
Figure~\ref{fig:U95-RAN63} illustrates the manifestation of linearity particularly well---spatial accuracy degrades significantly at the periphery, and error becomes especially high at the upper-right region.
Although this case that can likely be attributed to a sub-optimal device calibration or a poorly fitted headset, it is important to note that these conditions are inevitable in a real-world environment, and are therefore important to characterize. 
An analysis of linearity at this scale is useful for designing eye tracking applications that can accommodate this performance variation across the user's field of view.

\label{sec:discussion}
\section{Discussion}
\subsection{Spatial Accuracy and Precision}
We observe spatial accuracy metrics that are largely consistent with those produced in other VR headsets that have eye tracking functionality.
Our ``average'' spatial accuracy results---which we will designate as 1.08 dva, U50|E50 spatial accuracy in the RAN 127 task--- compare favorably to the median spatial accuracy values of 0.84 dva observed in the Vive Pro Eye~\cite{viveproeye} and 2.06 dva observed in the HoloLens 2~\cite{Aziz2022} in similar experimental conditions.
The spatial accuracy that we observe is additionally consistent with prior work in Meta Quest Pro, such as the analysis presented by~\cite{Wei2023}.
Our evaluation of spatial accuracy during the RAN127 task most closely resembles the conditions of~\cite{Wei2023} head-restrained experiments.
They report a median spatial accuracy of 2.015, which is comparable, if slightly worse, than our U50|E50 spatial accuracy of 1.08 dva.
Their median spatial precision of 0.632 dva is also comparable to our U50|E50 value of 0.66 dva.
We note that potential discrepancies in signal quality between studies can be caused by a number of difference between experimental protocols, such as the number of fixation targets (13 versus 49 targets), population size (12 versus 78 participants), the size of the targets (0.7 dva versus 0.57 dva), and differences in data processing strategies.

Eye tracking is emerging as a core functionality in VR platforms, and should therefore function effectively for a majority of users.
Designing for a wider population of users can have significant impacts on the overall layout of an gaze-based interface.
The spatial accuracy of an eye tracking system has a direct impact on the ideal size and orientation of interactable elements on a display~\cite{Feit2017}, the efficacy of dwell-based selection techniques~\cite{lohr2017comparison} and the overall reliability of applications that rely on accurate gaze estimation. 
We argue that it is not sufficient to design an eye tracking based application based on an ``average'' user's performance; in other words, allowing U50|E50 spatial accuracy metrics to dictate design decisions would produce an application that does not serve a sufficiently large proportion population. 
Instead, basing design decisions to accommodate users across the operational spectrum---say, all users who fall within the U75+ range---can produce eye tracking based applications that are accessible across a larger operational range.
For example, one may leverage the insights of the user- and error- percentile analysis presented in this paper to design interactable elements to provide functional interaction capabilities to the U75|E95 population, and create elements that are two to three times larger than the E50 results would suggest.
Measurements of extreme error values, such as those observed in the U95+|E95+ population tiers, can be considered outliers caused by factors such as occlusion and user noncompliance during task execution.
To avoid over-engineering for such noisy artifacts, we suggest that these extreme values may be disregarded for design purposes.



\subsection{Linearity}
Our analysis of linearity reveals a systematic relationship between spatial accuracy and target position across the entire population of participants.
Spatial accuracy in the horizontal direction tends to be better in the center of the field of view and worse toward the periphery.
Meanwhile, vertical spatial accuracy is typically better at the lower regions of the field of view, or may show no consistent relationship with target position at all.

In addition to observing a distinct correlation between target location and spatial accuracy across the entire population, we note that this correlation may be more pronounced in users at higher user percentile tiers, possibly due to a general decline in eye tracking signal quality.
To illustrate how linearity might manifest in individual users at different levels of operational performance, we provide 2-dimensional visualizations of spatial accuracy across the field of view for several users with different levels of signal quality (Figure~\ref{fig:2d-spatial-accuracy}). 
From these maps, it is clear that the degree to which spatial accuracy changes across the field of view can vary between users, and may additionally be influenced by the overall quality of eye tracking functionality for individual users.

While general descriptions of spatial accuracy are important for designing applications in VR, an analysis of linearity offers nuanced insights that further enhance the usability of eye tracking-based applications. 
For instance, one approach to accommodating the effect of linearity in user application design is to place critical interaction components in areas that consistently exhibit the best spatial accuracy~\cite{Komogortsev2011}.
Another approach could involve making interaction elements larger in areas with comparatively lower spatial accuracy.
Our detailed analysis of linearity offers a perspective on designing interfaces that not only facilitate basic functionality but also maximize the efficiency of gaze-based interaction.

\subsection{Luminance}
Our motivation for characterizing the effects of luminance in the Meta Quest Pro is to address the unique challenge that it presents in eye tracking devices embedded within VR platforms.
While an eye tracker's ability to capture gaze data does not solely depend on background illumination, our goal is to observe whether eye tracking performs well across a reasonably large range of luminance levels in our selected VR platform.
Because luminance is a controllable variable in VR platforms, it requires deliberate design choices to ensure that it does not interfere with eye tracking functionality.
Overall, we do not observe a substantial effect of background luminance on spatial accuracy or spatial precision. 
The distribution of spatial accuracy and spatial precision values across the user population do not change significantly between luminance conditions, which indicates that the Quest Pro is likely robust to the background luminance levels that we tested.
We observe that the nature of vertical linearity changes between the bright background and darker backgrounds, but it is not likely that luminance would have been the primary cause of this phenomenon.
Because details surrounding the Meta Quest Pro's gaze estimation mechanisms are not publicly available, we are unable to comment on how these findings reflect the inherent properties of the device's gaze estimation pipeline.
Nevertheless, it appears that the device's eye tracking performance is relatively similar across the two luminance levels that we employ in this study.



\subsection{Slippage}
Headset slippage can manifest differently across head-mounted eye tracking devices, but is often associated with a decline in eye tracking performance~\cite{Niehorster2020}.
Our headset slippage analysis shows that shifting the device on participants’ heads during the brow movement task clearly affects the performance of the eye tracking device.
When comparing performance before and immediately after the brow movement task was completed, we observed that spatial accuracy worsened as a result of this movement, but we observed that this degradation was not very large up to the 50th percentile of the user population.
Spatial accuracy significantly degrades for the 75th percentile of users and above, which can be problematic if not considered carefully during the design process.

We also measured spatial accuracy while users completed the brow task to assess the immediate impact of headset movement on eye tracking performance.
In this setting, we compared eye tracking signal quality during fixation period when participants' eyebrows were relaxed relative to when they were raised. Signal quality during the ``brows up'' condition was worse across the user population, but was far more pronounced in the U95+ subset of the population.

Although it appears from our presentation of Figure~\ref{fig:accuracy-slippage-intra-task} that the device exhibits only relatively low degradation in spatial accuracy overall, we must emphasize that this analysis was produced only after excluding 18 participants (23\% of the sample) because we failed to acquire data that was suitable for analysis. 
Failure-to-acquire rates are a valuable source of insight when evaluating a device for usability, as they can reveal situations in which the eye tracking functionality of the platform may become entirely unusable. 
Based on its high failure-to-acquire rate for the second slippage analysis, the device appears to be far less robust to headset slippage in the short term.
Our findings with the Meta Quest Pro underscore the continued need to embrace techniques that can mitigate the effect of headset slippage in wearable eye trackers, such as slippage-robust gaze estimation~\cite{Santini2019}.

We presented our battery of stimuli in the same order for every participant; participants completed the RAN 127 task, the BROW task, and then the RAN 63 task.
This sequence might have caused artifacts from headset slippage and fatigue to affect our analysis of luminance.


\section{Conclusion}
In this work, we explored the eye tracking signal quality of the Meta Quest Pro VR headset using a large dataset of eye movement recordings collected from 78 participants. 
In addition to presenting classical signal quality metrics--spatial accuracy, spatial precision, and linearity--in ideal settings, we also observe that external factors endemic to VR platforms such as background luminance and headset slippage can have a considerable effect on the eye tracking performance of this device. 
By expressing our evaluation of eye tracking signal quality through user percentiles across the population and error percentiles within individual users (i.e., our ``U,'' ``E'', and ``U|E'' metrics), we aim to encourage a user-centric approach to signal quality analysis that is informative for effectively designing eye tracking applications in VR.
We hope that our efforts will make effective design principles for usable eye tracking applications accessible to a wider audience, thereby advancing the adoption of eye tracking as a primary interaction modality in VR platforms.

\bibliographystyle{plain}
\bibliography{references}

\end{document}